\documentstyle[12pt]{article}
\parskip=\medskipamount

\def\appendix#1{
  \addtocounter{section}{1}
  \setcounter{equation}{0}
  \renewcommand{\thesection}{\Alph{section}}
 \section*{Appendix \thesection\protect\indent
 \parbox[t]{11.715cm} {#1}} 
 \addcontentsline{toc}{section}{Appendix \thesection\ \ \ #1}
  }

\renewcommand{\thefootnote}{\fnsymbol{footnote}}

\newcommand {\cA}{{\cal A}}
\newcommand {\cB}{{\cal B}}

\newcommand {\cD}{{\cal D}}

\newcommand {\cF}{{\cal F}}
\newcommand {\cG}{{\cal G}}

\newcommand {\cL}{{\cal L}}
\newcommand {\cM}{{\cal M}}
\newcommand {\cN}{{\cal N}}
\newcommand {\cO}{{\cal O}}
\newcommand {\cP}{{\cal P}}
\newcommand {\cQ}{{\cal Q}}

\newcommand {\cS}{{\cal S}}

\newcommand {\cU}{{\cal U}}

\newcommand {\cW}{{\cal W}}
\newcommand {\cX}{{\cal X}}

\newcommand {\cZ}{{\cal Z}}
\newcommand{\bA}{{\bf A}}
\newcommand{\bB}{{\bf B}}

\newcommand{\bR}{{\bf R}}

\newcommand{\bX}{{\bf X}}

\def\a{\alpha}
\def \bi{\bibitem}

\def\b{\beta}

\def\d{\delta}

\def\G{\Gamma}

\def\k{\kappa}
\def\l{\lambda}

\def\q{\theta}

\def\s{\sigma}

\def\F{\Phi}
\def\J{\Psi}
\def\L{\Lambda}
\def\O{\Omega}

\def\U{\Upsilon}

\newcommand{\ad}{{\dot{\alpha}}}                           
\newcommand{\bd}{{\dot{\beta}}}                            
\newcommand{\ve}{\varepsilon}                            

\newcommand{\pa}{\partial}                           
\newcommand{\hf}{\frac12}

\newcommand{\1}{\underline{1}}
\newcommand{\2}{\underline{2}}
\newcommand{\sect}[1]{\setcounter{equation}{0}\section{#1}}

\newcommand{\be}{\begin{equation}}
\newcommand{\ee}{\end{equation}}
\newcommand{\bea}{\begin{eqnarray}}
\newcommand{\eea}{\end{eqnarray}}
\newcommand{\non}{\nonumber}

\begin{document}

\begin{titlepage}
\thispagestyle{empty}

\begin{flushright}
LMU-TPW-00-03\\
hep-th/0001068 \\
January, 2000
\end{flushright}
\vspace{5mm}

\begin{center}
{\Large\bf  Supersymmetric Duality Rotations }\\
\end{center}

\begin{center} 
{\large 
Sergei M. Kuzenko\footnote{E-mail: 
sergei@theorie.physik.uni-muenchen.de} 
 and Stefan Theisen\footnote{E-mail: 
theisen@theorie.physik.uni-muenchen.de}
}\\
\vspace{2mm}

${}$\footnotesize{
{\it Sektion Physik, Universit\"at M\"unchen\\
Theresienstr. 37, D-80333 M\"unchen, Germany} 
} \\
\vspace{2mm}

\end{center}
\vspace{5mm}
                  
\begin{abstract}
\baselineskip=14pt
We derive $\cN = 1,~2$ superfield 
equations as the conditions 
for a (nonlinear) theory 
of one abelian $\cN=1$ or $\cN=2$ vector multiplet to be duality
invariant. The $\cN=1$ super Born-Infeld action is 
a particular solution of the corresponding equation.
A family of duality invariant nonlinear 
$\cN=1$ supersymmetric theories is described.
We present the solution of the $\cN=2$ duality equation
which reduces to the $\cN=1$ Born-Infeld action when the (0,1/2)
part of $\cN=2$ vector multiplet is switched off.
We also propose a constructive perturbative scheme to compute
duality invariant $\cN=2$ superconformal actions.
\end{abstract}
\vfill
\end{titlepage}

\newpage
\setcounter{page}{1}

\renewcommand{\thefootnote}{\arabic{footnote}}
\setcounter{footnote}{0}
\sect{Introduction}
The general theory of duality invariance 
of  abelian gauge theory was developed 
in \cite{GZ1,Z} and further elaborated in a series
of publications (see \cite{BinG,GR1,GR2,GZ2,GZ3,BMZ,Kam,ABMZ}
and references therein).
In this paper we generalize 
the duality equation of Gaillard and Zumino 
\cite{GZ2,GZ3}, 
also obtained independently in  \cite{GR1}, to
$\cN=1,~2$ supersymmetric theories. This 
duality equation is
the condition for a  theory
with Lagrangian $L(F_{ab})$ to be invariant under 
$U(1)$ duality transformations
\be
\d F ~=~ \l \, G~, \qquad \quad
\d G ~ =~ -\l \, F~, 
\ee
where
\be \tilde{G}_{ab} ~=~
\hf \, \ve_{abcd}\, G^{cd} ~=~ 
2 \, \frac{\pa L}{\pa F^{ab}}~.
\label{tilde-g}
\ee
The equation reads
\be
G^{ab}\, \tilde{G}_{ab} ~+~F^{ab} \, \tilde{F}_{ab} ~=~ 0
\label{GZ}
\ee
and presents a nontrivial constraint on the 
Lagrangian.

The Born-Infeld (BI) theory \cite{BI}
is a particular solution of eq. (\ref{GZ}).
The BI action naturally appears in string theory
\cite{FT,L}
(see \cite{T} for a recent review). 
Its $\cN=1$ supersymmetric 
generalization \cite{CF} (see also \cite{DP})
turns out to be the action for a Goldstone multiplet
associated with
partial breaking of $\cN=2$ to $\cN=1$ supersymmetry
\cite{BG,RT}. It has been conjectured \cite{BIK} that a
$\cN=2$ supersymmetric generalization of the
BI action should provide a model for
partial breakdown $\cN=4 \to \cN=2$, with the $\cN=2$
vector multiplet being the corresponding Goldstone field,
but the existing mechanisms of partial supersymmetry 
breaking are very difficult to implement 
in the $\cN=4$ case. A candidate for $\cN=2$
BI action has been suggested in \cite{Ket1}.
It correctly reduces to the Cecotti-Ferrara
action \cite{CF} once the $(0, \hf)$ part 
of the $\cN=2$ vector multiplet is switched off.
However, there exist 
infinitely many $\cN=2$ superfield actions
with that property.
Therefore, requiring the correct $\cN=1$ 
reduction does not suffice to fix a proper 
$\cN=2$ generalization of the BI action.  
One has to impose additional physical requirements.
Since no mechanism for partial 
$\cN=4 \to \cN=2$ breaking is currently available,
it is natural to look for the $\cN=2$ 
BI action as a solution of the supersymmetric generalization 
of the Gaillard-Zumino equation (\ref{GZ}).

In this paper we find $\cN=1,~2$ supersymmetric
generalizations of the duality equation (\ref{GZ}).
They are presented in eqs. (\ref{n=1dualeq})
and (\ref{n=2dualeq}), respectively.
It is not surprising that the Cecotti-Ferrara action 
\cite{CF} is a solution of the $\cN=1$ duality
equation. In contrast, the action proposed
in \cite{Ket1} does not satisfy the $\cN=2$ 
duality equation. However, the key to  
the construction of duality invariant $\cN=2$ BI action was 
given in \cite{Ket2} where a nonlinear $\cN=2$ 
superfield constraint was introduced as a minimal
extension of that generating the $\cN=1$ BI 
action \cite{BG,RT}. It was asserted
that the constrained superfield introduced does
generate the $\cN=2$ action given in \cite{Ket1}.
While this claim is incorrect, 
the constrained superfield nevertheless
does generate the duality invariant 
$\cN=2$ action that reduces to the $\cN=1$ BI action
after the $(0, \hf)$ part 
of the $\cN=2$ vector multiplet is switched off.

One application of
the $\cN=2$ duality equation may be 
to compute the duality invariant 
low-energy effective actions of
supersymmetric gauge theories. 
The $\cN=4$ super Yang-Mills theory is 
expected to be self-dual \cite{MO,Os}.
It was proposed in \cite{gkpr}
to look for its low-energy action on the Coulomb
branch as a solution of the self-duality equation
via the $\cN=2$ superfield Legendre transformation, and 
a few subleading corrections to the low-energy
action were determined. 
For non-supersymmetric theories it was shown in \cite{GZ3} that the  
Gaillard-Zumino equation (\ref{GZ}) 
implies self-duality via Legendre transformation.
The Gaillard-Zumino equation is much simpler
to solve and this advantage 
becomes essential in supersymmetric theories,
where the procedure of inverting the Legendre transformation
appears to be more involved at higher orders
of perturbation theory \cite{gkpr}.

We have already remarked that (\ref{GZ}) implies 
self-duality via Legendre transformation, but it is 
in fact a stronger condition. With reference to recent 
interest in the (supersymmetric) BI action within the 
context of D-branes, this stronger condition 
is in fact what one would like to impose. As was noted 
in \cite{D-branes,APPS,PR}, the D3-brane world-volume action, 
which contains, in addition to the gauge field also the 
axion and the dilaton fields, 
possesses a non-trivial $SL(2,{\bf R})$ symmetry. 
The BI action we are considering corresponds to the 
CP-even part of this action for the special choice of vanishing 
axion and dilaton. This background is invariant precisely under 
the $U(1)\subset SL(2,{\bf R})$ duality group we are considering.

Our paper is organized as follows.
In section 2 we derive the $\cN=1$ generalization 
of the Gaillard-Zumino equation and give
a family of duality invariant nonlinear $\cN=1$ models.
The $\cN=1$ BI action \cite{CF} is a special
member of this family. We also introduce a superconformally
invariant generalization of the $\cN=1$ BI action by 
coupling the vector multiplet to a scalar multiplet.
In section 3 we present the $\cN=2$ duality equation 
and derive its nonperturbative solution that reduces 
to the $\cN=1$ BI action when the $(0, \hf)$ part 
of $\cN=2$ vector multiplet is switched off.
We also develop a consistent perturbative scheme
of computing duality invariant $\cN=2$ superconformal
actions. 
In Appendix A we discuss the general structure of 
the duality equation in the non-supersymmetric case and
we show that any solution of 
(\ref{GZ}) admits a supersymmetric extension. 
In Appendix B we give an explicit proof that 
the $\cN=2$ BI action is self-dual with respect 
to Legendre transformation.

\sect{\mbox{$\cN$} = 1 duality rotations}

Let $S[W , {\bar W}]$ be the action describing the dynamics 
of a single $\cN=1$ vector multiplet. The (anti) chiral superfield
strengths ${\bar W}_\ad$ and $W_\a$,\footnote{Our $\cN=1$
conventions correspond to \cite{BK}.}  
\be
W_\a = -\frac{1}{4}\, {\bar D}^2 D_\a \, V~, \qquad \quad 
{\bar W}_\ad = -\frac{1}{4}\,D^2 {\bar D}_\ad \, V ~,
\ee
are defined in terms of a real unconstrained prepotential $V$. 
As a consequence, the strengths are constrained superfields,
that is they satisfy the Bianchi identity
\be
D^\a \, W_\a ~=~ {\bar D}_\ad \, {\bar W}^\ad~.
\label{n=1bi}
\ee

Suppose that $S[W , {\bar W}]$ can be unambiguously
defined\footnote{This is always possible if $S[W , {\bar W}]$ 
does not involve the combination $D^\a \,W_\a$ as an 
independent variable.}
 as a functional of {\it unconstrained}
(anti) chiral superfields ${\bar W}_\ad$ and $W_\a$.
Then, one can define (anti) chiral superfields
${\bar M}_\ad$ and $M_\a$  as
\be
{\rm i}\,M_\a \equiv 2\, \frac{\d }{\d W^\a}\,S[W , {\bar W}]
~, \qquad \quad
- {\rm i}\,{\bar M}^\ad \equiv 2\, 
\frac{\d }{\d {\bar W}_\ad}\, S[W , {\bar W}] ~.
\label{n=1vd}
\ee
The equation of motion following from the action $S[W,\bar{W}]$ reads
\be
D^\a \, M_\a ~=~ {\bar D}_\ad \, {\bar M}^\ad~.
\label{n=1em}
\ee

Since the Bianchi identity (\ref{n=1bi}) and the equation of
motion (\ref{n=1em}) have the same functional form,
one may consider infinitesimal $U(1)$ duality transformations
\be
\d W_\a ~=~ \l \, M_\a~, \qquad \quad
\d M_\a  ~=~ -\l \, W_\a~.
\label{n=1dt}
\ee
To preserve the definition (\ref{n=1vd}) of $M_\a$ and its 
conjugate, the action should transform as 
\be
\d S = -\frac{\rm i}{4}\,\l \,  \int {\rm d}^6 z\, 
\left\{W^\a W_\a - M^\a M_\a \right\} ~+~
\frac{\rm i}{4}\,\l \,  \int {\rm d}^6 {\bar z}\,
\left\{ {\bar W}_\ad{\bar W}^\ad  - 
{\bar M}_\ad {\bar M}^\ad \right\} ~,
\ee
in complete analogy with the analysis of \cite{GZ3}
for the non--supersymmetric case.\footnote{Note that the action 
$S$ itself is not duality invariant, 
but rather the combination
$S
-{{\rm i}\over 4}\int {\rm d}^6 z\, W M
+{{\rm i}\over4}\int {\rm d}^6 \bar z \,\bar W \bar M $.
The invariance of this functional under a finite $U(1)$
duality rotation by $  \pi / 2$, 
is equivalent to the self-duality
of $S$ under Legendre transformation, $S[W, \bar W ]
-{{\rm i}\over 2} \int {\rm d}^6 z\, W W_{\rm D}
+{{\rm i}\over2}\int {\rm d}^6 \bar z \,\bar W \bar W_{\rm D}
=S[W_{\rm D}, {\bar W}_{\rm D} ]\,$, with $W_{{\rm D}}$
being the dual chiral field strength.}
On the other hand, $S$ is a functional of $W_\a$ and 
${\bar W}_\ad$ only, and therefore it variations
under (\ref{n=1dt}) is
\be 
\d S = \frac{\rm i}{2}\,\l \, \int {\rm d}^6 z\, 
M^\a M_\a  ~-~
\frac{\rm i}{2}\,\l \,  \int {\rm d}^6 {\bar z}\, 
{\bar M}_\ad {\bar M}^\ad ~.
\ee
Since these two variations must coincide,
we arrive at the following reality condition
\be 
\int {\rm d}^6 z\, 
\Big( W^\a W_\a + M^\a M_\a \Big) ~=~
\int {\rm d}^6 {\bar z}\,
\Big( {\bar W}_\ad{\bar W}^\ad  +
{\bar M}_\ad {\bar M}^\ad \Big) ~.
\label{n=1dualeq}
\ee

In eq. (\ref{n=1dualeq}), the superfields
$M_\a$ and ${\bar M}_\ad$
are defined as in (\ref{n=1vd}), 
and $W_\a$ and ${\bar W}_\ad$ should be considered 
as {\it unconstrained} chiral and antichiral superfields,
respectively. Eq. (\ref{n=1dualeq}) is the condition 
for the $\cN=1$ supersymmetric theory to be 
duality invariant.   
We call it 
the $\cN=1$ duality equation.

A nontrivial solution of eq. (\ref{n=1dualeq}) is
the $\cN=1$ supersymmetric Born-Infeld
action \cite{CF,BG,RT} (see also \cite{DP})  
\bea
S_{\rm BI} &=& 
 \frac{1}{4}\int {\rm d}^6z \, W^2 +
\frac{1}{4}\int {\rm d}^6{\bar z} \,{\bar  W}^2 
+ { 1 \over g^4} \,  \int {\rm d}^8z \, \frac{W^2\,{\bar W}^2  }
{ 1 + \hf\, A \, + 
\sqrt{1 + A +\frac{1}{4} \,B^2} }~,
\label{bi} \\
 A &=&   { 1 \over 2g^4} \, 
\Big(D^2\,W^2 + {\bar D}^2\, {\bar W}^2 \Big)~,
\qquad
B = \, { 1 \over 2g^4} \, \Big(D^2\,W^2 - 
{\bar D}^2\, {\bar W}^2 \Big)~, \non 
\eea
where $g$ is a coupling constant. 
This is a model for a Goldstone multiplet 
associated with partial breaking of $\cN=2$ to $\cN=1$ 
supersymmetry \cite{BG,RT} (see also \cite{T}),
with $W_\a$ being the Goldstone multiplet.

New examples of $\cN=1$ duality invariant models
can be obtained by considering a 
general action of the form (see also Appendix A)
\be 
S ~=~ \frac{1}{4}\int {\rm d}^6z \, W^2 +
\frac{1}{4}\int {\rm d}^6{\bar z} \,{\bar  W}^2 
+  \frac{1}{2}\, \int {\rm d}^8z \, W^2\,{\bar W}^2  \,
L \big( D^2\,W^2 , {\bar D}^2\, {\bar W}^2 \big)~,
\label{gendualaction}
\ee
where $L(u,\bar u )$ is a real analytic function 
of the complex  variable $u \equiv D^2\,W^2$ and 
its conjugate. 
One finds 
\be
{\rm i}\,M_\a ~=~ W_\a \,\left\{\;
1 - \hf \,{\bar D}^2 \left[ {\bar W}^2 \left( L + 
D^2 \Big( W^2 \frac{\pa L}{\pa u} \Big) \right) \right]
\; \right\}~.
\ee
Then, eq. (\ref{n=1dualeq}) leads to 
\be
4\, \int {\rm d}^8z \, W^2\,{\bar W}^2  \,
\left( \G ~- ~{\bar \G} \right) ~=~
\int {\rm d}^8z \, W^2\,{\bar W}^2  \,
\left( \G^2 \, {\bar D}^2\, {\bar W}^2 
- {\bar \G}^2 \,D^2\,W^2  \right)~,
\ee
where 
\be
\G ~ \equiv ~ L ~+~
\frac{\pa L}{\pa u}\; D^2\,W^2 ~=~
\frac{ \pa (u\,L)}{\pa u} ~.
\ee
Since the latter functional relation must be satisfied
 for  arbitrary
(anti) chiral superfields ${\bar W}_\ad$ and
$W_\a$, we arrive at the following differential
equation for $L(u, \bar u )$:
\be
4\, \left( \frac{\pa (u\,L)}{\pa u}   ~- ~
\frac{\pa ({\bar u}\,L)}{\pa {\bar u}}   \right) ~=~
{\bar u}\, \left( 
\frac{\pa ( u\,L)}{\pa u} \right)^2
~-~ u \, \left( \frac{\pa ({\bar u}\,L)}{\pa {\bar u}}   
\right)^2  ~.
\label{dif}
\ee 
Similar to the non--supersymmetric case \cite{GR1,GZ3},
the general solution of this equation involves
an arbitrary real analytic function of a single real 
argument, 
$f({\bar u}\,u)$.\footnote{Among non--supersymmetric 
duality invariant models, 
only the Maxwell action 
and the BI action satisfy the requirement of 
shock-free wave propagation \cite{DMS}.}
It is an easy exercise to check that the $\cN=1$
BI action (\ref{bi}) satisfies eq. (\ref{dif}).

We conclude this section by giving an
extension of the model (\ref{bi}), in which
the vector multiplet is coupled to an external 
chiral superfield $\F$ in such a way that the system 
is not only duality invariant but also 
invariant under the $\cN=1$ superconformal group.
The action is
\bea
S &=& 
 \frac{1}{4}\int {\rm d}^6z \, W^2 +
\frac{1}{4}\int {\rm d}^6{\bar z} \,{\bar  W}^2 
+  \int {\rm d}^8z \, \frac{W^2\,{\bar W}^2 \,
(\F \,{\bar \F})^{-2} }
{1 + \hf\, \bA \, + 
\sqrt{1 + \bA +\frac{1}{4} \,\bB^2} \, 
}~,
\label{scbi} \\
 \bA &=&   { 1 \over 2} \, 
\left(\frac{ D^2\ }{{\bar \F}^2}\, 
\Big( \frac{W^2}{\F^2} \Big) 
+ \frac{{\bar D}^2\, }{\F^2} \, 
\Big( \frac{{\bar W}^2}{{\bar \F}^2}\Big) \right)~,
\qquad
\bB = \, { 1 \over 2} \, \left(
\frac{ D^2 }{{\bar \F}^2}\,  
\Big( \frac{W^2}{\F^2} \Big) - 
\frac{{\bar D}^2  }{\F^2} \, 
\Big( \frac{{\bar W}^2}{{\bar \F}^2}\Big) \right) ~. \non 
\eea
Superconformal invariance follows 
from the superconformal transformation 
properties as given in \cite{BKT}.
The theory is invariant under the duality rotations
(\ref{n=1dt}) with $\F$ being inert. 
By its very construction, the action is also invariant
under global phase transformations of $\F$. 
In a sense, this model is analogous to the BI theory
coupled to dilaton and axion fields \cite{GR2,BMZ}.

Similar to the analysis of \cite{BG,RT}, it is possible to show
that the action (\ref{scbi}) can be represented in the form
\be
S~=~  \frac{1}{4}\int {\rm d}^6z \, \bX +
\frac{1}{4}\int {\rm d}^6{\bar z} \,{\bar  \bX}~,
\ee
where the chiral superfield $\bX $ is a functional 
of $W_\a$ and ${\bar W}_\ad$ such that it satisfies
the nonlinear constraint
\be
\bX ~+~ \bX\,  \frac{ {\bar D}^2  }{4\F^2}\, 
\left(  \frac{{\bar \bX}}{{\bar \F}^2} \right) ~=~
W^2~.
\label{n=1constraint}
\ee  
The $\cN=1$ BI theory is obtained from this model
by freezing $\F$.

More generally, for any duality invariant system 
defined by eqs. (\ref{gendualaction}) and (\ref{dif}),
the replacement 
\be
W^2\,{\bar W}^2 ~ \longrightarrow ~
\frac{W^2\,{\bar W}^2}{ \F^2 \,{\bar \F}^2 }~,
\qquad D^2 ~ \longrightarrow ~
\frac{1}{ {\bar \F}^2 } \; D^2\;\frac{1}{ \F^2 } 
\ee
in (\ref{gendualaction}) preserves the duality invariance
but turns the action into a $\cN=1$ superconformal 
functional.

\sect{\mbox{$\cN$} = 2 duality rotations}
We now  generalize the results of the previous
section to the case of $\cN=2$ supersymmetry. 
We will work in $\cN =2$ global superspace 
${\bf  R}^{4|8}$ 
parametrized by 
$\cZ^A = (x^a, \q^\a_i, {\bar \q}^i_\ad) $,
where $i = {\1}, {\2}$. The flat covariant 
derivatives $\cD_A = (\pa_a, \cD^i_\a, 
{\bar \cD}^\ad_i )$ satisfy the standard algebra
\be
\{ \, \cD^i_\a ,\cD^j_\b \, \} =
\{ \, {\bar \cD}_{\ad i},
{\bar \cD}_{\bd j} \, \} =0~, \qquad
\{ \, \cD^i_\a , {\bar \cD}_{\ad j}\, \} =
-2\,{\rm i}\, \d^i_j \,(\s^a)_{\a \ad}\, \pa_a ~.
\ee
Throughout this section, we will use the
notation:
\bea
\cD^{ij} \equiv \cD^{\a (i} \cD^{j)}_\a 
= \cD^{\a i} \cD^{j}_\a  ~ , 
& \qquad &
{\bar \cD}^{ij} \equiv {\bar \cD}^{(i}_\ad {\bar \cD}^{j)\,\ad}
= {\bar \cD}^{i}_\ad {\bar \cD}^{j\,\ad} \non \\
 \cD^4 \equiv \frac{1}{16}
(\cD^{\underline{1}})^2\,
(\cD^{\underline{2}})^2~, & \qquad &
{\bar \cD}^4 \equiv \frac{1}{16} 
({\bar \cD}_{\underline{1}})^2\,
({\bar \cD}_{\underline{2}})^2  ~.
\eea
An integral over the full superspace can be reduce
to one over the chiral subspace
or over the antichiral subspace as follows:
\be
\int {\rm d}^{12}\cZ \; \cL (\cZ ) ~ =~ 
\int {\rm d}^{8}\cZ \; \cD^4\cL (\cZ )~=~
\int {\rm d}^{8}{\bar \cZ} \; {\bar \cD}^4\cL (\cZ )~.
\ee

\subsection{\mbox{$\cN$} = 2 duality equation}

The discussion in this subsection is completely analogous 
to the one presented in the first part of sect. 2. We will thus be brief. 
If $ \cS[\cW , {\bar \cW}]$ is the action describing the dynamics 
of a single $\cN=2$ vector multiplet, the (anti) chiral superfield
strengths ${\bar \cW}$ and $\cW$ are \cite{Mez}
\be
\cW =  {\bar \cD}^4 \cD^{ij} \, V_{ij}~, \qquad \quad 
{\bar \cW} = \cD^4 {\bar \cD}^{ij} \, V_{ij}
\ee
in terms of a real unconstrained prepotential $V_{(ij)}$. 
The strengths then 
satisfy the Bianchi identity \cite{gsw}
\be
\cD^{ij} \, \cW ~=~ 
{\bar \cD}^{ij} \, {\bar \cW}~.
\label{n=2bi-i}
\ee

Suppose that $\cS[\cW , {\bar \cW}]$ can be unambiguously
defined as a functional of {\it unconstrained}
(anti) chiral superfields ${\bar \cW}$ and $\cW$.
Then, one can define (anti) chiral superfields
${\bar \cM}$ and $\cM$ as
\be
{\rm i}\,\cM \equiv 4\, \frac{\d }{\d \cW}\,
\cS[\cW , {\bar \cW}]
~, \qquad \quad
- {\rm i}\,{\bar \cM} \equiv 4\, 
\frac{\d }{\d {\bar \cW}}\, \cS[\cW , {\bar \cW}]
\label{n=2vd}
\ee
in terms of which the equations of motion read
\be
\cD^{ij} \, \cM ~=~ {\bar \cD}^{ij} \, {\bar \cM}~.
\label{n=2em}
\ee

Again, since the Bianchi identity (\ref{n=2bi-i}) and the equation of
motion (\ref{n=2em}) have the same functional form,
one can consider infinitesimal $U(1)$ duality transformations
\be
\d \cW ~=~ \l \, \cM~, \qquad \quad
\d \cM  ~=~ -\l \, \cW~.
\label{n=2dt}
\ee
Repeating the analysis of Gaillard and Zumino \cite{GZ3}
(see also section 2), we now have to impose
\bea
\delta \cS &=& -{{\rm i}\over8}\, \l 
\int {\rm d}^8 \cZ \, \Big(\cW^2-\cM^2 \Big)
+{{\rm i} \over 8} \, \l
\int {\rm d}^8\bar{\cZ}\, 
\Big(\bar \cW^2-\bar \cM^2 \Big) \\
&=& {{\rm i}\over4} \, \l \int {\rm d}^8\cZ\,\cM^2
-{{\rm i} \over4} \, \l \int {\rm d}^8\bar\cZ\,\bar \cM^2 \non
\eea
The theory is thus duality invariant
provided the following reality condition is satisfied:
\be 
\int {\rm d}^8 \cZ\, 
\Big( \cW^2 + \cM^2 \Big) ~=~
\int {\rm d}^8 {\bar \cZ}\,
\Big( {\bar \cW}^2  +
{\bar \cM}^2 \Big)\,. 
\label{n=2dualeq}
\ee
Here $\cM$ and ${\bar \cM}$
are defined as in (\ref{n=2vd}), 
and $\cW$ and ${\bar \cW}$ should be considered 
as {\it unconstrained} chiral and antichiral superfields,
respectively. 
Eq. (\ref{n=2dualeq}) serves as our master functional equation
to determine duality invariant models of the 
$\cN=2$ vector multiplet. We remark that, as in the $\cN=1$ case, 
the action itself is not duality invariant, but 
\be
\delta\, \Bigg(\cS-{{\rm i}\over8}\int {\rm d}^8\cZ\, \cM\cW
+{{\rm i}\over 8}\int {\rm d}^8\bar\cZ\,\bar \cM\bar \cW
\Bigg) ~=~0~.
\ee
The invariance of the latter functional under a finite $U(1)$
duality rotation by $  \pi / 2$, 
is equivalent to the self-duality
of $\cS$ under Legendre transformation, 
\be
\cS[\cW, {\bar \cW} ]
- {{\rm i}\over 4} \int {\rm d}^8 \cZ\, \cW \cW_{\rm D}
+ {{\rm i}\over 4}\int {\rm d}^8 {\bar \cZ} \,
{\bar \cW} {\bar \cW}_{\rm D}
~=~\cS[\cW_{\rm D}, {\bar \cW}_{\rm D} ]~,
\ee 
where the dual chiral field strength
$\cW_{{\rm D}}$ is given by eq. (\ref{dfs}).

\subsection{\mbox{$\cN$} = 2 BI action}

Recently, Ketov \cite{Ket1} suggested the following 
action 
\bea
\cS_{\rm BI} &=& 
 \frac{1}{8}\int {\rm d}^8 \cZ \, \cW^2 +
\frac{1}{8}\int {\rm d}^8{\bar \cZ} \,{\bar  \cW}^2 
+ { 1 \over 4} \,  \int {\rm d}^{12} \cZ \, 
\frac{\cW^2\,{\bar \cW}^2  }
{ 1 - \hf\, \cA \, + 
\sqrt{1 - \cA +\frac{1}{4} \,\cB^2} }~,
\label{kbi} \\
 \cA &=& \cD^4\,\cW^2 + {\bar \cD}^2\, {\bar \cW}^2 ~,
\qquad
\cB = \cD^4\,\cW^2 - 
{\bar \cD}^4\, {\bar \cW}^2  \non 
\eea
as the $\cN=2$ supersymmetric generalization 
of the BI action. We will first demonstrate that it indeed reduces
to the $\cN=1$ BI action. We then show that this condition is 
not strong enough to uniquely fix the $\cN=2$ BI action but this 
is possible if, in addition, one imposes eq. (\ref{n=2dualeq}).  

Given a $\cN=2$ superfield $U$, its $\cN=1$ projection 
is defined to be $ U| = U(\cZ)|_{ \q_{ \underline{2} } = 
{\bar \q}^{{\2}} = 0}$. The $\cN=2$ vector multiplet
contains two independent chiral $\cN=1$ components
\be 
\cW | = \sqrt{2}\, \F~, \qquad \quad \cD_\a^{\underline{2}}\, 
\cW |= 2{\rm i}\, W_\a~,\qquad\quad(\cD^{\underline{2}})^2\cW|
=\sqrt{2} \, \bar{D}^2\bar\F~.
\ee
Using in addition that
\be
\int{\rm d}^8\cZ=-{1\over4}\int{\rm d}^6 z \, 
(\cD^{\underline{2}})^2~,
\qquad
\int{\rm d}^{12}\cZ={1\over16}\int{\rm d}^8 z \,
(\cD^{\underline{2}})^2 \, (\bar\cD_{\underline{2}})^2 ~,
\ee
the free $\cN=2$ vector multiplet action
straightforwardly reduces to $\cN=1$ superfields 
\be
\cS_{\rm free} = 
\frac{1}{8}\int {\rm d}^8 \cZ \, \cW^2 +
\frac{1}{8}\int {\rm d}^8{\bar \cZ} \,{\bar  \cW}^2 
=  \int {\rm d}^8z \, {\bar \F} \F +
 \frac{1}{4}\int {\rm d}^6z \, W^2 +
\frac{1}{4}\int {\rm d}^6{\bar z} \,{\bar  W}^2~. 
\ee

If one switches off $\F$,
\be
\F = 0~ \qquad \Longrightarrow \qquad 
( \cD^{\underline{2}} )^2 \cW| = 0~,
\label{f=0}
\ee
the action (\ref{kbi}) reduces to the $\cN=1$
BI theory (\ref{bi}) (with $g=1$).
However, as we will now demonstrate, 
there exist infinitely many 
$\cN=2$ actions with that property.\footnote{
The property $W_\a \,W_\b \, W_{\gamma} =0$
of the $\cN=1$ vector multiplet, which is crucial in the 
discussion of the $\cN=1$ BI action, has no direct analog for
its $\cN=2$ counterpart.} 
To demonstrate why this is possible, 
consider the following obviously  
different functionals
\bea
&& \int {\rm d}^{12} \cZ \,
\cW^2 {\bar \cW}^2 \left\{ \,
(\cD^4 \cW^2)^2 {\bar \cD}^4 {\bar \cW}^2  
+ ({\bar \cD}^4 {\bar \cW}^2)^2 \cD^4 \cW^2 \,  \right\}~, 
\non \\
&& \int {\rm d}^{12} \cZ \,
\cW^2 {\bar \cW}^2 \left\{ \,
( \cD^4 \cW^2 )\, {\bar \cD}^4 \Big[{\bar \cW}^2
\cD^4 \cW^2 \Big] + ({\bar \cD}^4 {\bar \cW}^2)\,
\cD^4 \Big[\cW^2 {\bar \cD}^4 {\bar \cW}^2 \Big]
\, \right\}\,. \non
\eea
They coincide under (\ref{f=0}).
Therefore, the requirement of correct $\cN=1$ reduction
is too weak to fix a proper $\cN=2$ generalization 
of the BI action\footnote{It was claimed in \cite{Ket1,Ket2}
that the action (\ref{kbi}) is self-dual with respect to
the $\cN=2$ Legendre transformation. This is, however, not correct.}.

We suggest to search for a $\cN=2$ generalization of the 
BI action as a solution of the $\cN=2$ duality equation
(\ref{n=2dualeq}) compatible with the requirement 
to give the correct 
$\cN=1$ reduction. We have checked to some order in 
perturbation theory that these two requirements
uniquely fix the solution:
\bea
\cS_{\rm BI} &=& 
\frac{1}{8}\int {\rm d}^8 \cZ \, \cW^2 +
\frac{1}{8}\int {\rm d}^8{\bar \cZ} \,{\bar  \cW}^2 
~+~ \cS_{\rm int} ~, \non \\
\cS_{\rm int} &=& 
{ 1 \over 8} \,  \int {\rm d}^{12} \cZ \, 
\cW^2\,{\bar \cW}^2\, \Bigg\{ 1 + 
\hf \, \Big( \cD^4 \cW^2 + {\bar \cD}^4 {\bar \cW}^2 \Big) 
\label{per} \\
&+& \frac{1}{4} \, 
\Big( (\cD^4 \cW^2)^2 + ({\bar \cD}^4 {\bar \cW}^2)^2 \Big)
+ \frac{3}{4}\, (\cD^4 \cW^2)({\bar \cD}^4 {\bar \cW}^2)
\non \\
&+&  \frac{1}{8}\, 
\Big( (\cD^4 \cW^2)^3 + ({\bar \cD}^4 {\bar \cW}^2)^3 \Big)
\non \\
&+& \hf \,\Big( (\cD^4 \cW^2)^2 ({\bar \cD}^4 {\bar \cW}^2)
+ (\cD^4 \cW^2) ({\bar \cD}^4 {\bar \cW}^2)^2 \Big) \non \\
&+& \frac{1}{4} \Big(
( \cD^4 \cW^2 )\, {\bar \cD}^4 \Big[{\bar \cW}^2
\cD^4 \cW^2 \Big] + ({\bar \cD}^4 {\bar \cW}^2)\,
\cD^4 \Big[\cW^2 {\bar \cD}^4 {\bar \cW}^2 \Big] \Big)
\Bigg\} ~~+~~ O(\cW^{12})~. \non
\eea
The expression in the last two lines of (\ref{per})
constitutes the leading perturbative corrections
where our solution of the duality equation (\ref{n=2dualeq})
differs from the action (\ref{kbi}).

We now present the nonperturbative
solution of (\ref{n=2dualeq})
which reduces to the $\cN=1$ BI action (\ref{bi})
under the condition (\ref{f=0}). The action reads
\be
\cS_{\rm BI} =
\frac{1}{4}\int {\rm d}^8 \cZ \, \cX +
\frac{1}{4}\int {\rm d}^8{\bar \cZ} \,{\bar \cX}~,
\label{n=2bi}
\ee
where the chiral superfield $\cX$ is a functional 
of $\cW$ and $\bar \cW$ defined via the 
constraint\footnote{The property $\bX^2 =0$
of the $\cN=1$ constraint (\ref{n=1constraint})
has no direct analog for $\cX$.}
\be
\cX ~=~ \cX \, {\bar \cD}^4 {\bar \cX} ~+~
\hf \, \cW^2~.
\label{n=2con}
\ee
Solving it iteratively for $\cX$ one may 
verify the equivalence of (\ref{n=2bi})
and (\ref{per}) up to the indicated order.
The constraint (\ref{n=2con}) was introduced
in \cite{Ket2} as a $\cN=2$ generalization of that 
generating the $\cN=1$ BI action (\ref{bi})
\cite{BG,RT} (see eq. (\ref{n=1constraint})). 
It was also claimed in \cite{Ket2}
that the action (\ref{kbi})  can be equivalently
described by eqs. (\ref{n=2bi}) and (\ref{n=2con}).
This is clearly incorrect, since they 
lead to the action (\ref{per}) rather than to (\ref{kbi}).
But the constraint (\ref{n=2con})
has a deep origin: 
the $SL(2,\bR)$ invariant system introduced in \cite{BMZ}
admits a minimal $\cN=2$ extension on the base
of the constraint (\ref{n=2con}) such that the original
$SL(2,\bR)$ invariance remains intact.

Let us prove that the system described by eqs. 
(\ref{n=2bi}) and (\ref{n=2con}) provides
a solution of the duality equation (\ref{n=2dualeq}).
Under an infinitesimal variation of $\cW$ only, we have 
\bea
\d_{\cW}  \cX &=& \d_{\cW}  \cX \,
{\bar \cD}^4 {\bar \cX} + \cX\,{\bar \cD}^4  
\d_{\cW}  {\bar \cX} + \cW \, \d \cW~, \non \\
\d_{\cW}  {\bar \cX} &=& \d_{\cW}  {\bar \cX} \,
\cD^4 \cX + {\bar \cX}\, \cD^4 \d_{\cW} \cX~.
\eea
{}From these relations one gets
\be
\d_{\cW}  \cX = \frac{1}{1 - \cQ }\,
\left[ \frac{ \cW \, \d \cW}
{ 1 - {\bar \cD}^4 {\bar \cX} } \right] ~, \qquad
\d_{\cW}  {\bar \cX} =
\frac{\bar \cX}{1 - \cD^4 \cX } \, \cD^4 \d_{\cW} \cX~,
\ee
where
\bea
 \cQ = \cP \, {\bar \cP}~, & \qquad &
{\bar \cQ} = {\bar \cP}\, \cP~, \non \\
\cP = \frac{\cX}{1 - {\bar \cD}^4 {\bar \cX} }\,
{\bar \cD}^4~, & \qquad &
{\bar \cP} = \frac{\bar \cX}{1 - \cD^4 \cX } \, \cD^4~.
\eea
With these results, it is easy to compute $\cM$:
\be
{\rm i}\, \cM = \frac{\cW}{1 - {\bar \cD}^4 {\bar \cX} }\,
 \left\{ 1 + {\bar \cD}^4\, {\bar \cP}\, \frac{1}{1 - \cQ }\,
\frac{\cX}{1 - {\bar \cD}^4 {\bar \cX} } +
{\bar \cD}^4\, \frac{1}{1 - {\bar \cQ} }\,
\frac{\bar \cX}{1 - \cD^4 \cX } \right\}~.
\label{M}
\ee 
Now, a short calculation gives
\be
{\rm Im}\,\int {\rm d}^8 \cZ \,\left\{
 \cM^2 
+2\, \frac{1}{1 - \cQ }\,
\frac{\cX}{1 - {\bar \cD}^4 {\bar \cX} } 
\right\} ~=~ 0~.
\label{prom}
\ee
On the other hand, the constraint (\ref{n=2con}) implies
\be
\int {\rm d}^8 \cZ \, \cX - 
\int {\rm d}^8{\bar \cZ} \,{\bar \cX} =
\hf \, \int {\rm d}^8 \cZ \, \cW^2 -   
\hf \,\int {\rm d}^8{\bar \cZ} \,{\bar \cW}^2 ~,
\label{real}
\ee
and hence
\be
\frac{\d }{\d \cW}\,\left\{ \int {\rm d}^8 \cZ \, \cX - 
\int {\rm d}^8{\bar \cZ} \,{\bar \cX} \right\} = \cW~.
\ee
The latter relation can be shown to be equivalent to 
\be
 \frac{1}{1 - \cQ }\,
\frac{\cX}{1 - {\bar \cD}^4 {\bar \cX} }
= \cP \,\frac{1}{1 - {\bar \cQ} }\, 
\frac{\bar \cX}{1 - \cD^4 \cX } + \cX~.
\label{importantrel}
\ee 
Using this result in eq. (\ref{prom}), 
we arrive at the relation
\be
\int {\rm d}^8 \cZ \, \cM^2 - 
\int {\rm d}^8{\bar \cZ} \,{\bar \cM}^2 =
-2 \int {\rm d}^8 \cZ \, \cX
+2 \int {\rm d}^8{\bar \cZ} \,{\bar \cX}
\ee 
which is equivalent, due to (\ref{real}),
to  (\ref{n=2dualeq}).

In Appendix B we
prove the self-duality of the $\cN=2$ BI action under
Legendre transformation explicitly, 
although this property already follows from 
the general analysis of \cite{GZ3}
or our discussion in subsect. 3.1.

\subsection{Duality invariant
\mbox{$\cN$} = 2 superconformal actions}

The $\cN=4$ super Yang-Mills theory is believed 
to be self-dual \cite{MO,Os}. It was therefore 
suggested in \cite{gkpr}
to look for its low-energy effective action on the Coulomb branch
as a solution to the self-duality equation via the $\cN=2$
Legendre transformation such that the leading 
(second- and fourth- order) terms in the momentum
expansion of the action look like
\be
\cS_{\rm lead} ~=~ \frac{1}{8}
\int {\rm d}^8 \cZ \, \cW^2 ~+~
\frac{1}{8}\int {\rm d}^8{\bar \cZ} \,{\bar  \cW}^2 
~+~ \frac{1}{4}\, c \int {\rm d}^{12} \cZ \,   
\ln {\cW}\, \ln {\bar \cW} ~+~ \dots~,
\label{leading}
\ee 
where the third term represents the leading quantum
correction computed in \cite{F-to-the-four,gkpr}. 

In our opinion, the perturbative scheme of solving 
the self-duality equation via the $\cN=2$
Legendre transformation is difficult \cite{gkpr}
as one has to invert the Legendre transformation.
We suggest to look for the low-energy action 
of $\cN=4$ SYM as a solution of the $\cN=2$ duality
equation (\ref{n=2dualeq}). This equation
is easy to deal with and it
implies self-duality via Legendre transformation. 

The low-energy effective action we are looking for 
should be in addition invariant under the $\cN=2$ 
superconformal group. This means that, along 
with the structures given in (\ref{leading}),
the action may involve the following manifestly 
superconformal functionals \cite{BKT}
\bea
\cS_1 &=& \int {\rm d}^{12} \cZ \,
\ln \cW \;
\L ( \nabla
 \ln \cW  )\, 
  ~+~{\rm c.c.} ~, \label{s2} \\
\cS_2 &=& \int {\rm d}^{12} \cZ \, 
\U( \nabla \ln \cW \, ,\, {\bar \nabla} \ln {\bar \cW}  ) ~,
\label{s3}
\eea
where 
\be
\nabla ~ \equiv ~ \frac{1}{ {\bar \cW}^2}\, 
\cD^4 ~, \qquad 
{\bar \nabla} ~ \equiv ~ 
\frac{1}{ \cW^2}\, 
{\bar \cD}^4~,
\label{sc-in-vop}
\ee
and $\L$ and $\U$ are arbitrary holomorphic and real 
analytic functions, respectively. 
The superfields $\nabla \ln \cW$  
and ${\bar \nabla} \ln {\bar \cW}$ prove to be 
superconformal scalars \cite{BKT}. 
The main property of the operators
(\ref{sc-in-vop}) is that,
for any superconformal scalar $\J$,
$\nabla \J$ and 
${\bar \nabla} \J$ are also superconformal scalars.

In components, the functionals (\ref{leading}),
(\ref{s2}) and (\ref{s3}) contain
all possible structures which involve the physical 
scalar fields $\varphi = \cW|_{\q = 0}$ and the
electromagnetic field strength $F_{ab}$
(where $F_{\a \b} \propto \cD_\a{}^i \cD_{\b \,i} \cW|_{\q = 0}$)
without derivatives, along with terms containing derivatives
and auxiliary fields. 
Simple power counting
determines the necessary number of covariant derivatives in the 
action in order to produce a given power of $F$.
Since $F \propto \cD^2 \cW$, there should be $4n$ 
$\cD$'s in the superfield Lagrangian to get 
$F^{4 +2n}$  (additional 8 derivatives come from 
the superspace measure, $\int {\rm d}^{12} \cZ  =
\int {\rm d}^4 x \, \cD^4 {\bar \cD}^4$).

We are looking for a perturbative solution
of (\ref{n=2dualeq}) in the framework 
of the momentum expansion
or, equivalently, as a series in powers
of $\nabla$ and $\bar \nabla$.
But with the Ansatz $\cS = \cS_{\rm lead} +\cS_1 +
\cS_2$ it is easy to see that no solution 
of (\ref{n=2dualeq}) exists. To obtain a consistent 
perturbation theory, we should allow for 
higher derivatives. More precisely, we should 
add new terms such that any number of operators
$\nabla$ and $\bar \nabla$ are inserted 
in the Taylor expansion of $\U$ (\ref{s3}).
In other words, $\cS_2$ should be extended to a more
general functional $\hat{\cS}_2$
which can be symbolically written as
\footnote{There exist more general superconformal 
invariants of the $\cN=2$ vector multiplet \cite{BKT},
as compared to the action (\ref{tilde-s}), 
and some of them were determined in \cite{gkpr}
from the requirement of scale and $U(1)_R$
invariance.
It suffices for our purposes that (\ref{tilde-s})
provides a consistent Ansatz to solve the $\cN=2$ 
duality equation (\ref{n=2dualeq}).} 
\be 
\hat{\cS}_2 ~=~ \int {\rm d}^{12} \cZ \, 
\hat{ \U } ( \nabla \ln \cW \, ,\, 
{\bar \nabla} \ln {\bar \cW} \, ,\,
\nabla \, ,\, \bar \nabla   ) ~.
\label{tilde-s}
\ee
{}For the action 
\be
\cS[ \cW, {\bar \cW}] ~=~ \cS_{\rm lead} +\cS_1 +
\hat{\cS}_2
\ee
the equation of motion can be represented in terms of
\be
{\rm i} \, \cM ~\equiv~
4\, \frac{\d }{\d \cW}\,
\cS[\cW , {\bar \cW}]
~=~ \cW\, \Big\{\, 1 ~+~
{\bar \nabla} \G \, \Big\}~,
\ee
for some functional $\G(  \ln \cW  , 
 \ln {\bar \cW} ,
\nabla , \bar \nabla   )$ such that  
$\G = c \ln {\bar \cW} +
O(\nabla  )$. Then, the duality equation 
(\ref{n=2dualeq}) is equivalent to 
\be
{\rm Im} \int {\rm d}^{12} \cZ \, 
\left\{ 2\,\G ~+~ \G\, {\bar \nabla}\, \G \right\}~= ~0~.
\label{working}
\ee

In the framework of perturbation theory,
the procedure of solving of eq. (\ref{working}) amounts
to simple algebraic operations.
To low order in the perturbation theory, 
the solution reads
\bea
\cS &=& \frac{1}{8}
\int {\rm d}^8 \cZ \, \cW^2 +
\frac{1}{8}\int {\rm d}^8{\bar \cZ} \,{\bar  \cW}^2
+ \frac{1}{4}\int {\rm d}^8{\bar \cZ} \,\cL ~,\non \\
\cL &=& c \, \ln {\cW}\, \ln {\bar \cW}
+ \frac{1}{4}\, c^2\, \Big( 
\ln \cW \,\nabla \ln \cW ~~+~~ {\rm c.c.} \Big) \non \\
&&+ \frac{1}{4}\, c^3 \,d\, 
( \nabla \ln \cW ) \, {\bar \nabla}\ln {\bar \cW}
- \frac{1}{8}\, c^3 \, 
\Big( 
\ln \cW \, (\nabla \ln \cW )^2 
~~+~~ {\rm c.c.} \Big) \non \\
&&+   \frac{1}{16}\,  c^4\, \Bigg(
 (1-4d)\, ( \nabla \ln \cW )^2\, {\bar \nabla} \ln {\bar \cW}
+(2d-1)\, ( \nabla \ln \cW ) \, 
{\bar \nabla} \nabla  \ln \cW \non \\
&&  \qquad \qquad  \qquad
+ \frac{5}{3} \, \ln \cW \, (\nabla \ln \cW )^3 
~~+~~ {\rm c.c.} \Bigg)
~~+~~ O(\nabla^4)~. 
\label{sdaction}
\eea 
Here $d$ is the first parameter in the derivative 
expansion of $\cS$ which is not fixed by the $\cN=2$ duality 
equation (\ref{n=2dualeq}). 
Note that if we had only imposed the condition of self-duality 
under Legendre transformation, as was done in \cite{gkpr}, 
we could not have fixed the coefficent $-{1\over8}c^3$ 
of the fourth term in ${\cal L}$. 
In general, for 
any self-conjugate monomial
in the expansion of $\cS$, like 
$( \nabla \ln \cW ) \, {\bar \nabla}\ln {\bar \cW}$,
the corresponding coefficient is not determined by
eq. (\ref{n=2dualeq}) in  terms of those appearing
in the structures in $\cS$ with less derivatives.
However, such coefficients can be fixed if one imposes
some additional conditions on the solution of 
eq. (\ref{n=2dualeq}). For example, one can require 
the solution to reduce to a given $\cN=1$ action under 
the condition $\cW | = {\rm const}$.

It should be pointed out that the $c^3$--corrections
in (\ref{sdaction}) have been determined in \cite{gkpr}
by solving 
the self-duality equation via the $\cN=2$
Legendre transformation. 
To compute the ${\cal O}(c^4)$ term via the duality equation
(\ref{n=2dualeq}) involves only 
elementary algebraic manipulations.

As is seen from (\ref{sdaction}), solutions
of the duality equation (\ref{n=2dualeq})
contain higher derivative structures 
${\bar \nabla} \nabla  \ln \cW$, 
$\nabla {\bar \nabla} \nabla  \ln \cW$, etc.
What is the fate of such terms?
The striking result of \cite{gkpr}
is the fact that, to the order $c^3$, there 
exists a nonlinear $\cN=1$ superfield redefinition 
which eliminates
all higher derivative (accelerating) 
component structures (contained 
already in the first term of $\cL$ (\ref{sdaction})).
The price for such a redefinition is that 
the original linear $\cN=2$ supersymmetry 
turns into a nonlinear one being typical
for D3-brane actions \cite{APS}.
The nonlinear redefinition of \cite{gkpr}
eliminates the higher derivative terms 
to some order of perturbation theory, 
but it in turn generates new  such terms 
at higher orders in the momentum expansion.
Therefore, in order for such a nonlinear redefinition 
to be consistently defined, 
the superfield action should involve higher derivatives
of arbitrary order. The duality equation
(\ref{n=2dualeq}) might guarantee the existence of a
consistent redefinition to eliminate acceleration 
terms.

\vskip.5cm

\noindent
{\bf Acknowledgements} \hfill\break
We are grateful to Gleb Arutyunov, Evgeny Ivanov
and Arkady Tseytlin for helpful discussions. 
Support from DFG, the German National Science Foundation,
from GIF, the German-Israeli foundation 
for Scientific Research and from the EEC under TMR
contract ERBFMRX-CT96-0045 is gratefully acknowledged.
This work was also supported in part
by the NATO collaborative research grant PST.CLG 974965,
by the RFBR grant No. 99-02-16617, by the INTAS grant
No. 96-0308 and  by the DFG-RFBR grant No. 99-02-04022.

\setcounter{section}{0}
\setcounter{subsection}{0}

\appendix{
\mbox{$\cN=0$} duality invariant models}

In this appendix we give several equivalent forms
of the Gallard-Zumino equation (\ref{GZ})
by representing the Lagrangian $L(F_{ab})$ as a
real function of one complex variable, 
\bea
& L(F_{ab})= L(\cU , \bar{\cU} )~, \qquad \quad
\cU = \cF + {\rm i} \, \cG~, & \non \\
& \cF = \frac{1}{4} \, F^{ab} F_{ab}~, \qquad \quad
\cG = \frac{1}{4} \, F^{ab} \tilde{F}_{ab} ~. &
\eea
The theory is parity invariant iff $L(\cU , \bar{\cU} )
= L( \bar{\cU}, \cU )$. 

One calculates $\tilde{G}$ (\ref{tilde-g}) to be 
\be
\tilde{G}_{ab} = \Big( F_{ab} +{\rm i}\, \tilde{F}_{ab}\Big)
\, \frac{\pa L}{\pa \cU} ~+~
\left( F_{ab} - {\rm i} \, \tilde{F}_{ab} \right)\, 
\frac{\pa L}{\pa \bar{\cU}}~,
\ee
and the Gallard-Zumino equation (\ref{GZ}) takes the form
\be 
{\rm Im}\;  \left\{ \cU - 4\, \cU\, 
\left( \frac{\pa L}{\pa \cU} \right)^2 \right\} = 0~,
\label{GZ2}
\ee
which is equivalent to the equations 
obtained in \cite{GR1,GZ3} but turns out to be more 
convenient for supersymmetric generalizations.
If one splits $L$ into the sum of Maxwell's part 
and an interaction, 
\be
L = -\hf \, \Big( \cU + \bar{\cU} \Big) ~+~
L_{{\rm in}}~, \qquad \quad 
L_{{\rm in}} = \cO (| \cU |^2)~,
\ee
the above equation turns into 
\be 
{\rm Im}\;  \left\{ \cU \, \frac{\pa L_{{\rm in}} }{\pa \cU}
-  \cU\, 
\left( \frac{\pa L_{{\rm in}}   }{\pa \cU} \right)^2 \right\} = 0~.
\label{GZ3}
\ee
We restrict $L_{{\rm in}}$ to be a real analytic function of
$\cU$ and $\bar \cU$. Then, every solution of eq. (\ref{GZ3})
is of the form
\be
L_{{\rm in}} (\cU, \bar{\cU} ) ~=~
\cU \, \bar{\cU} \; \O (\cU, \bar{\cU} )~,
\qquad \quad \O = \cO (1)~,
\ee
where $\O$ satisfies  
\be
{\rm Im}\;  \left\{ \frac{\pa (\cU \, \O) }{\pa \cU}
- \bar{\cU}\, 
\left( \frac{\pa (\cU \, \O )  }{\pa \cU} \right)^2 \right\} = 0~.
\label{GZ4}
\ee
Note that for any solution 
$L_{{\rm in}} (\cU, \bar{\cU} )$ of 
(\ref{GZ3}), or any solution $\O (\cU, \bar{\cU} )$  
of (\ref{GZ4}), the functions
\be 
\hat{L}_{{\rm in}} (\cU, \bar{\cU} ) =
\frac{1}{\k^2} \, L_{{\rm in}} (\k^2 \,\cU, 
\k^2 \, \bar{\cU} )~, \qquad
\hat{\O} (\cU, \bar{\cU} ) = \k^2 \,
\O (\k^2 \,\cU, \k^2 \, \bar{\cU} )
\ee
are also solutions of eqs. (\ref{GZ3}) and (\ref{GZ4}),
respectively, for arbitrary real parameter $\k^2$.

Up to a trivial rescaling, eq. (\ref{GZ4}) coincides
with the $\cN=1$ duality equation (\ref{dif}).
Therefore, any non-supersymmetric duality invariant 
model admits a $\cN=1$ supersymmetric extension given
by eqs. (\ref{gendualaction}) and (\ref{dif}). 
It is easy to read off the bosonic sector of the 
action (\ref{gendualaction}). 
{}For vanishing fermionic fields, 
$W_\a|_{\q = 0} = 0$, one finds \cite{BK} 
\be
\frac{1}{8} \, D^2 \, W^2|_{\q = 0} 
= \cU - 2 \, \cD^2 ~,
\ee
where $\cD (x)$ is the auxiliary field of the $\cN=1$ 
vector multiplet. If we take the solution $\cD=0$ 
of the equation of motion for $\cD$, then the action
(\ref{gendualaction}) reduces to a generic $\cN=0$ 
duality invariant model.

\appendix{
\mbox{$\cN=2$} BI action and Legendre \\ transformation}

To prove that the system defined by eqs.
(\ref{n=2bi}) and (\ref{n=2con}) is self-dual 
under Legendre transformation, we replace the action 
(\ref{n=2bi}) by the following one
\be
\cS=
\frac{1}{4}\int {\rm d}^8 \cZ \, \bigg\{
\cX [\cW, {\bar \cW}] - {\rm i}\,\cW\,\cW_{\rm D} \bigg\}
+
\frac{1}{4}\int {\rm d}^8{\bar \cZ} \,
\bigg\{   {\bar \cX}[\cW, {\bar \cW}] + 
{\rm i}\, {\bar \cW} {\bar \cW}_{\rm D} \bigg\} ~,
\label{a1}
\ee
where $\cW$ is now considered to be an unconstrained
chiral superfield, and its dual chiral strength 
$\cW_{\rm D}$ reads
\be
\cW_{\rm D} =  {\bar \cD}^4 \cD^{ij} \, U_{ij}~, 
\label{dfs}
\ee
with $U_{ij}$ an unconstrained real prepotential.
The equation of motion for $U_{ij}$ implies the Bianchi 
identity (\ref{n=2bi-i}), and hence the action reduces to 
(\ref{n=2bi}). On the other hand, varying the action with respect 
to $\cW$ leads to
\be
\cW_{\rm D} ~=~ \cM~,
\label{dualstrength}
\ee
where $\cM$ is given in eq. (\ref{M}).
The latter equation can be solved to express 
$\cW$ in terms of $\cW_{\rm D}$ and its conjugate.
Instead of doing this explicitly, we note  that 
eqs. (\ref{importantrel}) and (\ref{dualstrength})
allow one to rewrite the action (\ref{a1}) as
\be
\cS=
\frac{1}{4}\int {\rm d}^8 \cZ \, 
\cX_{\rm D} +
\frac{1}{4}\int {\rm d}^8{\bar \cZ} \,
{\bar \cX}_{\rm D} ~,
\ee
where 
\be 
\cX_{\rm D} ~ \equiv ~ -
 \frac{1}{1 - \cQ }\,
\frac{\cX}{1 - {\bar \cD}^4 {\bar \cX} }
- \cP \,\frac{1}{1 - {\bar \cQ} }\, 
\frac{\bar \cX}{1 - \cD^4 \cX } ~.
\ee 
Using eqs. (\ref{importantrel}) and (\ref{dualstrength})
once more, one can prove that $\cX_{\rm D}$
satisfies the constraint
\be
\cX_{\rm D} ~=~ \cX_{\rm D} \, {\bar \cD}^4 
{\bar \cX}_{\rm D} ~+~
\hf \, \cW_{\rm D}{}^2~.
\ee
This completes the proof.

\end{document}